 \documentclass[preprint,12pt]{elsarticle}

\usepackage{amssymb}
\usepackage{comment}

\journal{Nuclear Physics A}

\def\ffc{F_{\rm C}}
\def\ffm{F_{\rm M}}
\def\het{^3{\rm He}}

\def\hbq{{\hat Q}}
\def\hq{{\hat q}}
\def\hp{{\hat p}}
\def\hbp{{\hat P}}
\def\hk{{\hat k}}
\def\hko{{\hat k}_1}
\def\hkd{{\hat k}_2}
\def\hkt{{\hat k}_3}
\def\hkop{{\hat k}'_1}
\def\lb{{\rm L}}
\def\seta{\sqrt{\eta}}
\def\seto{\sqrt{1+\eta}}
\def\bq{\mathbf{q}}
\def\bp{\mathbf{p}}
\def\bbq{\mathbf{Q}}

\begin{document}

\begin{frontmatter}



\title{Relativistic rank-one separable kernel for helium-3 charge form factor}

\author[label1]{Serge  Bondarenko}\ead{bondarenko@jinr.ru}
\author[label1]{Valery Burov}
\author[label1]{Sergey Yurev}
\address[label1]{BLTP, Joint Institute for Nuclear Research, Dubna, 141980, Russia }

\begin{abstract}
Elastic electron-helium-3 scattering is studied in the relativistic impulse
approximation. The amplitudes for the three-nucleon system
are obtained by solving the relativistic generalization of the Faddeev
equations with a separable rank-one kernel of the nucleon-nucleon interaction.
The contribution of partial-wave states with nonzero orbital moment
of a nucleon pair is investigated.
The static approximation and additional
relativistic corrections for the $\het$ charge form factor are calculated
for the momentum transfer squared up to 100 fm$^{-2}$.
The contribution of the $^3D_1$ partial-wave state
and relativistic corrections are found to be sizable at high energies.
\end{abstract}




\end{frontmatter}



\section{Introduction}
\label{intro}

The study of electron-nucleus reactions is important for obtaining information
on the nucleon-nucleon (NN) interaction, which is crucial for understanding
the structure of strong interactions. At present, a lot of experimental data
are known for the reaction cross sections and polarization observables.
Of great interest is the intermediate range of energies in these reactions
where the nonrelativistic description based on the potential or
nonrelativistic meson-nucleon models does not properly work. On the other hand,
Quantum Chromodynamics, which operates with quark and gluon degrees of freedom,
also does not give an appropriate description in this region.

The latest experimental data for elastic electron-deuteron/helium
scattering (for example~\cite{Arnold:1978qs,Camsonne:2016ged}, for $\het$ elastic
form factors at high momentum transfer squared up to 60-100 fm$^{-2}$),
require taking into account the relativistic treatment of nuclear systems.

One of the promising relativistic approaches is the covariant formalism
based on the Bethe-Salpeter equation for two nucleons~\cite{Salpeter:1951sz}.
This approach operates with relativistic meson-nucleon degrees of freedom.
{ There are a lot of investigations of two-nucleon systems and their
reactions~(\cite{Tjon:1979ih,Zuilhof:1980ae,Hummel:1989qn,Hummel:1990zz,Hummel:1993fq},
see also the comprehensive review~\cite{Marcucci:2015rca}).}

Calculations with covariant separable kernels of NN interaction are also
widely used. The results of the investigations for the elastic electron-deuteron scattering
can be found
in~\cite{Rupp:1989sg,Rupp:1985ic,Bondarenko:2000yg,Bondarenko:2002zz,Bekzhanov:2013yoa,Bekzhanov:2014kxa}.

Using the NN interaction to calculate three-nucleon (3N) systems may shed light
on the occurrence of non-nucleonic degrees of freedom or effective 3N
forces.
The Faddeev equations~\cite{Faddeev:1960su}
are usually used~(see also~\cite{Sitenko:1971})
to study three-particle systems in quantum mechanics. The equations describe
these systems with pair potentials of any kind as bound and scattering states.
{
The relativistic generalization
of the Faddeev equations can be
applied to a system of three relativistic particles~\cite{Rupp:1987cw,Rupp:1991th,Bondarenko:2015kma,Bondarenko:2017ibt,Bondarenko:2018xoq}.}
In this case, the relativistic two-particle T~matrix is taken as a solution
of the Bethe-Salpeter equation (we call such a three-particle equation the
Bethe-Salpeter-Faddeev equation, BSF).

{
  Among the relativistic methods for the three-nucleon system, one should also note the relativistic
  approach, which uses the covariant spectator theory (CST)
  to construct the three-nucleon wave function~{\cite{Stadler:1996ut,Stadler:1997iu}}
  and the variational Monte Carlo calculations with the relativistic Hamiltonian~\cite{Carlson:1993zz}.
}

{
One of the important physical problems
is to give the right location of the first diffraction minimum of the three-nucleon charge form factor.
In the review~\cite{Marcucci:2015rca}, three different relativistic theoretical approaches
are considered for the form factors of the two- and three-nucleon systems:
one is based on realistic interactions and currents, including relativistic corrections,
the other relies on a chiral effective field theory description of strong
interactions in nuclei and the last one is CST.
For momentum transfers squared below 25 fm$^{-2}$ there is satisfactory
agreement between experimental data and theoretical results in all three approaches.
However, above 25 fm$^{-2}$ a relativistic treatment of the dynamics is necessary.
}

To solve the BSF equation, it is necessary to know the potential of the
nucleon-nucleon interaction in the explicit form.
The separable kernels of NN interactions can be
used~\cite{Rupp:1987cw,Rupp:1991th,Bondarenko:2015kma,Bondarenko:2017ibt,Bondarenko:2018xoq}
to simplify the calculations.
In recent works, we have solved the BSF equation for rank-one separable
kernels of NN interactions and took into account the two-nucleon states
with the total angular momentum $j=0-1$~\cite{Bondarenko:2017mef}. For the sake
of simplicity, we considered the nucleon propagators in the scalar-particle
form while the spin-isospin dependence was treated by applying the recoupling
matrix~\cite{Bondarenko:2018xoq}. The form factors of the potential were the relativistic
Yamaguchi functions~\cite{Yamaguchi:1954mp,Yamaguchi:1954zz}.

{
One of the problems of the numerical solution of the BSF equations
is that the solution is usually obtained in Euclidean space,
while the calculation of the electromagnetic (EM) reaction observables
requires a relativistic wave function in Minkowski space.
It should be noted that there are papers on such a difficult
task~\cite{Ydrefors:2019jvu,Ydrefors:2020duk}.
Another way is to carry out the Taylor series expansion of the
corresponding arguments.
}

{ Relativistic calculations for elastic electron scattering off $^3$He
were considered in several papers~\cite{Gross:2003qi,Pinto:2009dh,Pinto:2009jz,Rupp:1987cw,Rupp:1991th}.}
In~\cite{Bondarenko:2019gcd}, we studied the sensitivity of unpolarized
elastic electron scattering off $^3$He to nucleon EM
form factors in the static approximation.

In the present paper, we considered two types of additional effects:
the contribution of partial-wave states with nonzero orbital moment
of a nucleon pair and 
the relativistic corrections for the $\het$ charge form factor.
The latter are due to the full Lorentz transformation of the four-momentum
of the one-nucleon propagator, the residue at a simple pole on the complex
variable $q_0$ and the terms of a Taylor series expansion
of the arguments of the outgoing particle wave function.

The paper is organized as follows: in Sec. 2, the expressions for the charge
$^3$He form factor are given, in Sec. 3 the static approximation and
relativistic corrections are defined, in Sec. 4 the calculations and results
are discussed and finally the conclusion is given.

\section{Form factors of a three-nucleon system}
As a system with one-half spin, the electromagnetic current of $^3$He
can be parameterized by two elastic form factors: charge (electric) $\ffc$ and
magnetic $\ffm$ (see for example,~\cite{Sitenko:1971}).
In calculations, we use a straightforward relativistic generalization
of the nonrelativistic expression for the $^3$He charge form factor, which
has the following form~\cite{Schiff:1964zz,Gibson:1965zza,Sitenko:1971,Rupp:1987cw}:
\begin{eqnarray}
&&2\ffc =  (2F^p_{\rm C} + F^n_{\rm C} )F_1  - \frac23(F^p_{\rm C} - F^n_{\rm C})F_2 +  2(F^n_{\rm C} - F^p_{\rm C})F_3, 
\label{F_He_ch}
\\
\nonumber
\end{eqnarray}
where $F^{p,n}_{\rm C}$ is the charge form factor of the proton and neutron,
respectively.

The functions $F_{1,2,3}$ can be expressed in terms of the wave
functions of the three-nucleon system. In the relativistic impulse approximation,
they can be written as follows:
\begin{eqnarray}
F_{1}(\hbq) =  \int d^4{\hp} \int d^4{\hq}\, G_1'(\hkop)\, G_1(\hko)\, G_2(\hkd)\, G_3(\hkt)\,
\sum_{i=1}^3\Psi^*_i(\hp,\hq;\hbp)\, \Psi_i(\hp,\hq';\hbp'),
\nonumber\\
F_{2}(\hbq) = -3 \int d^4 {\hp} \int d^4 {\hq}\, G_1'(\hkop)\, G_1(\hko)\, G_2(\hkd)\, G_3(\hkt)\,
\Psi^*_1(\hp,\hq;\hbp)\, \Psi_2(\hp,\hq';\hbp'),
\nonumber\\
F_{3}(\hbq) = \int d^4 {\hp} \int d^4 {\hq}\, G_1'(\hkop)\, G_1(\hko)\, G_2(\hkd)\, G_3(\hkt)\,
\Psi^*_3(\hp,\hq;\hbp)\, \Psi_4(\hp,\hq';\hbp'),
\label{f123}
\end{eqnarray}
where $\hbq$ is the four-momentum transfer, $\hp$ and $\hq, \hq'=\hq-\frac23\hbq$ are
the Jacobi four-momenta of the three-nucleon system, and $\hbp,\hbp'=\hbp+\hbq$ are the total four-momenta.
The nucleon propagators read:
\begin{eqnarray}
  &&  G_1(\hk_1) =  \left[(\frac13 \hbp - \hq)^2  - m_N^2 + i\epsilon\right]^{-1},
  \label{g123}\\
  &&  G_1'(\hk_1') = \left[(\frac13 \hbp' - \hq')^2  - m_N^2 + i\epsilon\right]^{-1},
  \nonumber\\
  &&  G_2(\hk_2) = \left[(\frac13 \hbp + \frac12 \hq + \hp)^2 - m_N^2 + i\epsilon\right]^{-1},
  \nonumber\\  
  &&  G_3(\hk_3) = \left[(\frac13 \hbp + \frac12 \hq - \hp)^2 - m_N^2 + i\epsilon\right]^{-1},
\nonumber
\end{eqnarray}
where $m_N$ is the nucleon mass and the infinitesimal negative addition to nucleon mass
defines how to treat propagator singularities.

The three-nucleon wave functions $\Psi_i$ can be
expressed
in terms of the following functions:
\begin{eqnarray}
&&\Psi_1 = A(1) + A(2) + A(3),   \label{4func1}\\
&&\Psi_2 = \frac12(B(2) + B(3) -2B(1))  +   \frac{\sqrt 3}{2}(C(3) - C(2)),\nonumber\\
&&\Psi_3 = \frac{\sqrt{3}}{2}(B(3) - B(2)) - \frac12(C(3) + C(2) -2C(1)), \nonumber\\
&&\Psi_4 = D(1) + D(2) + D(3),\nonumber
\end{eqnarray} 
which are related to the spin-isospin $a(s,i)$ states of the two-particle
subsystem:
\begin{eqnarray}  
&& A(n)  = \frac1{\sqrt{2}}(u_3(n) -  u_1(n)),  \label{4func2}\\
&& B(n)  = \frac1{\sqrt{2}}(u_3(n) +  u_1(n)),  \nonumber\\
&& C(n)  = \frac1{\sqrt{2}}(u_2(n) -  u_4(n)),  \nonumber\\
&& D(n)  = -\frac1{\sqrt{2}}(u_2(n) +  u_4(n))\nonumber
\end{eqnarray}
where $n$ is the number of a particle.

The functions $u_a$ correspond to the following two-nucleon sub-states:
\begin{eqnarray}
&& a=1:(s,i) = (1,0) [^3S_1,^3D_1],\\
&& a=2:(s,i) = (1,1) [^3P_0,^3P_1],\nonumber\\
&& a=3:(s,i) = (0,1) [^1S_0],\nonumber\\
&& a=4:(s,i) = (0,0) [^1P_1].\nonumber
\end{eqnarray}

We consider various combinations of the spin, isospin
and momenta of the three nucleons under their permutations,
to satisfy the Pauli principle (the $^3$He vertex function must be antisymmetric
with respect to permutation of any pair of particles)
to obtain above-mentioned equations~(see details in~\cite{Schiff:1964zz,Gibson:1965zza}).

The wave functions $u_a$ can be related to the vertex functions, which are
the BSF equation solutions $\Phi_a$. The relations
in the center-of-mass (c.m.) frame of the three-nucleon system
$\hbp_{c.m.}=(\sqrt{s},\mathbf{0})$ read:
\begin{eqnarray}
&&  u_a(\hp,\hq) = \sum_{l \lambda L M_L}\, {\cal Y}_{l \lambda L M} (\mathbf{n_p},\mathbf{n_q})\,
  g_l^a(p_0,p)\, \tau_l^{a}[(\frac23\sqrt{s}+q_0)^2-\bq^2]\, \Phi^a_{l \lambda L}(q_0,q),\nonumber\\
&&  {\cal Y}_{l \lambda L M_L} (\mathbf{n_p},\mathbf{n_q}) =
      \sum_{m \mu} (l m \lambda \mu|LM_L) {Y}_{l m}(\mathbf{n_p}) {Y}_{\lambda \mu}(\mathbf{n_q}).
\label{ua}
\end{eqnarray}

The form of (\ref{ua}) is a consequence of using the kernel of NN interaction
in a separable form. Here $(.|.)$ is the Clebsch-Gordan coefficient,
${Y}_{l m}(\mathbf{n})$ are
spherical harmonics, $l$ and $m$ are the two-particle
subsystem angular momentum and its projection,
$\lambda$ and $\mu$ are the angular momentum of the third particle relative
to the two-particle subsystem and its projection, $L$ and $M_L$ are the total
angular momentum and its projection, $g_l^a$ are the form factors
of the NN interaction kernel and $\tau_l^{a}$ is the relativistic
two-particle propagator of interacting nucleons.

In the previous paper~\cite{Bondarenko:2019gcd}, the calculations were carried out
using only the $^1S_0,^3S_1$ partial-wave states.
In this paper, the $D$ and $P$ partial-wave states are added and their contributions are considered.

To calculate the functions $F_{1,2,3}$ (\ref{f123}), it is convenient to use the Breit
reference system that is defined as follows:
\begin{eqnarray}
  \hbq=(0,\bbq),\qquad \hbp = (E_B,-\frac{\bbq}{2}),\qquad \hbp' = (E_B,\frac{\bbq}{2}),
  \label{brt}
\end{eqnarray}
with $E_B = \sqrt{{\bbq}^2/4+s}$.
Also, $\sqrt{s}=M_t=3m_N-E_t$ and $E_t$ are the mass and the binding energy
of a three-nucleon bound state, respectively.

The solutions of the BSF equations $\Psi_i$ in (\ref{f123}), however,
have been found in the c.m. (rest) frames of the corresponding three-nucleon systems.
To relate the Breit and initial particle c.m. frames, the Lorentz transformation $\lb$
should be applied to the four-momenta:
\begin{eqnarray}
  \hbp = \lb \hbp_{c.m.},\qquad \hp = \lb \hp_{c.m.},\qquad \hq = \lb \hq_{c.m.}.
  \label{lbp}
\end{eqnarray}
The explicit form of the transformation $\lb$ can be obtained by using~(\ref{brt}).
Let us assume the boost of the system to be along the $Z$ axis:
\begin{eqnarray}
  \lb =
  \left(
  \begin{array}{cccc}
    \seto & 0 & 0 & -\seta \\
    0 & 1 & 0 & 0 \\
    0 & 0 & 1 & 0 \\    
    -\seta & 0 & 0 & \seto
  \end{array}
  \right).
\end{eqnarray}  
Here the parameter $\eta = \bbq^2/4s$ is introduced.
It is clear from~(\ref{brt}) that the Breit and final particle c.m. frames
are related as follows:
\begin{eqnarray}
  \hbp' = \lb^{-1} \hbp'_{c.m.},\qquad \hp' = \lb^{-1} \hp'_{c.m.},\qquad \hq' = \lb^{-1} \hq'_{c.m.}.
  \label{lbpp}
\end{eqnarray}

Since the four-momenta $\hq$ and $\hq'=\hq-\frac23\hbq$ are defined in different c.m. frames,
the relations between their components are rather lengthy (we omit the c.m. labels):
\begin{eqnarray}
  &&q_{0}'  =  (1 + 2\eta)\, q_0 - 2\seta \seto\, q_z + \frac23\seta\,Q,
  \label{lzt}\\
  &&q'_x =  q_x \qquad q'_y = q_y  \nonumber\\
  &&q'_z = (1 + 2\eta)\, q_z - 2\seta\seto\,q_0 -\frac23 \seto\, Q,
\nonumber
\end{eqnarray}
here $q_z = q y$ is the projection of momentum $\bq$ onto the $Z$ axis and $y=\cos{\theta_{qQ}}$.

To summarize, the arguments of the initial and final particle wave functions and propagators
were expressed in terms of the momenta calculated in the corresponding c.m. frames and
related to each other using the Lorentz transformation. The Jacobian of the boost $\lb$
is equal to one, so integration in~(\ref{f123}) is performed on the $p$ and $q$ variables
in the initial particle c.m. frame.

\section{Static approximation and relativistic corrections}

Using the results of the previous section, one can write for the propagators
\begin{eqnarray}
  &&  G_1(q_0,q) =  \left[(\frac13 \sqrt s - q_0 )^2  -  \bq^2 -m_N^2+i\epsilon\right]^{-1} ,
  \label{g123p}\\
  &&  G_1'(q_0',q') = \left[(\frac13 \sqrt s - q'_0 )^2  - \bq'^2  -m_N^2+i\epsilon\right]^{-1},
  \nonumber\\
  &&  G_2(p_0,p,q_0,q) =
  \left[(\frac13 \sqrt s + (\frac12q_0 + p_0 ))^2  - \bp^2 -  \frac14\bq^2 -  {\bp} \cdot {\bq} -m_N^2+i\epsilon\right]^{-1},
  \nonumber\\  
  &&  G_3(p_0,p,q_0,q) = 
  \left[(\frac13 \sqrt s + (\frac12q_0 - p_0 ))^2  - \bp^2 -  \frac14\bq^2 +  {\bp} \cdot {\bq} -m_N^2+i\epsilon\right]^{-1},
\nonumber
\end{eqnarray}
with ${\bp} \cdot {\bq} = pqx$ and $x=\cos{\theta_{pq}}$.
The arguments of the final particle wave function are related to the initial ones
by expressions~(\ref{lzt}).

Since the solutions of the BSF equations have been obtained in the Euclidean space~\cite{Bondarenko:2018xoq}
and are known only for real values of $q_4$, the simplest way to
calculate~(\ref{f123}) is to apply the Wick rotation procedure $p_0 \to ip_4, q_0 \to iq_4$,
if it is possible.
Thus, one needs to investigate the
analytic structure of the integrand on the complex-valued variables $p_0,q_0$.
The location of the variable $p_0$ singularities allows one to apply the Wick
rotation procedure, although this is not true
for the variable $q_0$ in general.

\subsection{Static approximation}

First, the so-called static approximation (SA) is considered.
The SA assumes that all terms in the Lorentz transformation~(\ref{lzt})
proportional to $\eta$ are canceled:
\begin{eqnarray}
q_{0}'  =  q_0, \qquad {\bf q'} = {\bf q} -\frac23 {\bf Q},\label{lzt0}
\end{eqnarray}
and these expressions are put as arguments into the propagator $G_1'(k_1')$ and
final particle wave function $\Psi(p,q')$:
\begin{eqnarray}
  && G_1'(q_0',q') \to
  \left[(\frac13 \sqrt s - q_0 )^2  -  \bq^2 - \frac23 {\bq \cdot \bbq} - \frac49 \bbq^2 - m_N^2 +i\epsilon\right]^{-1},
  \label{sa}\\
  && \Psi_i(p_0,p,q_0',q') \to \Psi_i(p_0,p,q_0,|{\bf q} -\frac23 {\bf Q}|),
  \nonumber
\end{eqnarray}
with ${\bq \cdot \bbq} = qQy$.

Analyzing~(\ref{sa}), one can see that the poles of $G_1'$ on $q_0$ do not cross
the imaginary $q_0$ axis and always stay in the second and fourth quadrants.
In this case, the Wick rotation procedure $q_0 \to iq_4$ can be applied.

{
The static approximation is similar to nonrelativistic calculations in that they use
arguments (\ref{lzt0}) of the wave functions. However, the main difference is the dependence on
$q_0$ of the wave functions and propagators that have positive and negative energy parts.
}

\subsection{Relativistic corrections}

Now we discuss the relativistic corrections (RC) to the SA.
Second, the full expressions~(\ref{lzt}) for $q'$ are used in $G_1'(k_1')$
and (\ref{lzt0}) in $\Psi(p,q')$:
\begin{eqnarray}
&&  G_1' = \label{lb}\\
&&  \left[q_0^{2} + \frac23\sqrt{s}(1 + 6\eta) q_0 
       + 4 \seto  \sqrt{s}\seta q_z - \frac83 \eta s
   + \frac19 s   - {\bq}^2 - m_N^2 + i\epsilon\right]^{-1},
  \nonumber\\
&&  \Psi_i(p_0,p,q_0',q') \to \Psi_i(p_0,p,q_0,|{\bf q} -\frac23 {\bf Q}|).
  \nonumber
\end{eqnarray}

{
Analyzing~(\ref{lb}), one can see that for any
$t = -\hbq^2 > -\hbq_{min}^2 = 2/3\sqrt{s}(3m_N-\sqrt{s})$
the pole of $G_1'$ on $q_0$ crosses the imaginary $q_0$ axis and
appears in the third quadrant.
}

In this case, using the Cauchy theorem, one can transform
the integrals over $p_0$, $q_0$
as follows:
{
\begin{eqnarray}
  &&
  \int_{-\infty}^{\infty}dp_0 \int_{-\infty}^{\infty} dq_0\,
  \int_{0}^{\infty} dq\,\int_{-1}^{1} dy\, ...\, f(p_0,q_0,p,q,x,y)= 
  \label{intpol}\\
  &&
  -  \int_{-\infty}^{\infty}dp_4 \int_{-\infty}^{\infty} dq_4\,
  \int_{0}^{\infty} dq\,\int_{-1}^{1} dy\, ...\, f(ip_4,iq_4,p,q,x,y)
  \nonumber \\
  &&  
  + 2\pi \mathop{\mathrm{Res}}_{q_0 = q_{0}^{(2)} } \int_{-\infty}^{\infty}dp_4\,
  \int_{q_{min}}^{q_{max}} dq\,\int_{y_{min}}^{1} dy\, ...\, f(ip_4,q_0,p,q,x,y),
  \nonumber
\end{eqnarray}
where (...) means the two-fold integral $\int_{0}^{\infty} dp\,\int_{-1}^{1} dx$ and}
\begin{eqnarray}
q_0^{(1,2)} = \frac{\sqrt{s}}{3}(1+6\eta)
\pm \sqrt{4\eta(1+\eta)s -4\sqrt{s}\sqrt{\eta} \sqrt{1+\eta} q y  +\bq^2 + m_N^2}
\end{eqnarray}
are the simple poles of the propagator $G'_1$.

The first integral on the right-hand side of~(\ref{intpol})
is a six-fold integral.
The second one is a five-fold integral with
the limits of integration on $q$ and $y$
\begin{eqnarray}
&& q_{min,max} = 2\sqrt{s}\seta\seto \mp \frac13 \sqrt{s + 12\eta s + 36\eta^2s - 9m_N^2},\\
&& y_{min} = \frac{1}{36}\frac{24\eta s + 9m_N^2 + 9\bq^2 - s}{q\sqrt{s}\seta\seto},\qquad\qquad y_{max} = 1,
\end{eqnarray}
and the residue at the point $q_0=q_0^{(2)}$ is calculated.
Remembering that the BSF solutions are known for real values of $q_4$ only,
the following assumption was made:
\begin{eqnarray}
  \Psi(p_0,p,q_0',q') \to g(p_0,p)\,\tau[(\frac23\sqrt{s}+q_0^{(2)})^2-{\bar \bq}'^2]\, \Phi(0,{\bar q}'),
  \nonumber
\end{eqnarray}
where value ${\bar q}'$ is obtained using~(\ref{lzt}) with $q_0 = q_0^{(2)}$.

Thus, two new effects appear: the Lorentz boost in the $G_1'(k_1')$
arguments, which gives a boost contribution (BC),
and a simple pole on $q_0$, which gives an additional term
in integrals -- a pole contribution (PC).

Third, one can take into account the Lorentz boost of the arguments
of the final $\het$ wave function.
As was already mentioned, since the solutions of the BSF equations were obtained
in the Euclidean space and are known only for real values of $q_4$, it is impossible to
take into account the full Lorentz transformation of $q'$. However, one can
carry out a Taylor series expansion contribution (EC) around the point~(\ref{lzt0})
on the parameter $\eta$.

The expansion of the function $\Phi(q'_4,q')$ up to the first order of the parameter
$\eta$ has the following form:
\begin{eqnarray}
\Phi(iq'_4,q') =  \Phi(iq_4,|\mathbf{q} - \frac23\mathbf{Q}|)
+\Big[ C_{q_4} \frac{\partial}{\partial q_4} \Phi_j(iq_4,q) \Big]_{q=|\mathbf{q} - \frac23\mathbf{Q}|}
\nonumber\\
+\Big[ C_{q} \frac{\partial}{\partial q}\Phi_j(iq_4,q) \Big]_{q=|\mathbf{q} - \frac23\mathbf{Q}|},
\label{df}
\end{eqnarray}
where
\begin{eqnarray}
&&C_{q_4} = -i\left(2i\eta q_4  - 2 \seta\seto q y+ \frac23 \seta Q\right),
\nonumber\\
&&C_{q} = \left( 2\eta q y - 2i \seta\seto
q_4   -  \frac23 (\seto - 1)Q \right) y.
\nonumber
\end{eqnarray}
To exclude double counting, one needs to take into account only the second and third terms
of the right-hand expression of~(\ref{df}) since the first term coincides with BC.

The derivative of the vertex function  $\Phi$ can be found by differentiating
the integral equation  $\Phi = \int K \Phi$.
In this case, the function $\Phi'$ is determined by the integral
$\Phi' = \int K' \Phi$ where $K'$
is a derivative of the kernel of the integral equation.

\section{Calculations and results}

To calculate the functions $F_{1,2,3}$, the analytic expressions for $g^a(p)$
and $\tau^{a}(s,q)$ were used. The numerical solution for the functions
$\Phi_a(q)$ was obtained by solving the system of homogeneous integral BSF 
equations by means of the Gaussian quadratures (see details
in~\cite{Bondarenko:2015kma}) and then interpolated to the $(q_4,q)$ points
of integration. The Vegas algorithm of the Monte-Carlo integrator
was used to perform multiple integration in equations~(\ref{f123}).
The dipole fit was used for nucleon form factors.

\begin{figure}[ht]
  \centering
  \includegraphics[width=0.85\linewidth,angle=0]{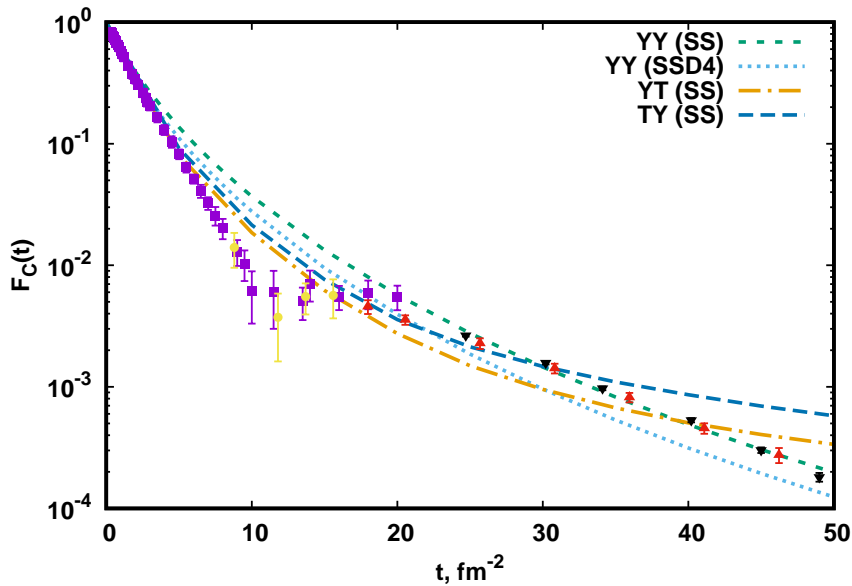}
  \caption{Static approximation for the charge form factor of $^3$He
    as a function of $t$
    for the rank-one separable Yamaguchi and Tabakin potentials.
    The experimental data
    from~\cite{Mccarthy:1977vd} are denoted by boxes,
    from~\cite{Bernheim:1972az} are denoted by circles,
    from~\cite{Arnold:1978qs} are denoted by triangles-up,
    from~~\cite{Camsonne:2016ged} are denoted by triangles-down.}
  \label{fig1}
\end{figure}

Figure~\ref{fig1} shows the static approximation for different rank-one separable
Yamaguchi and Tabakin potentials
for the $\het$ charge form factor as a function of the momentum transfer squared $t$.
The line denoted as YY (SS) stands for the $^1S_0$ and $^3S_1$ partial-wave states calculated with the
Yamaguchi potential,
the lines denoted as YT (SS) and TY (SS) stand for the $^1S_0$ and $^3S_1$ partial-wave states calculated
with the Yamaguchi (Y) and Tabakin (T) potentials, respectively.
The combinations of the Yamaguchi and Tabakin potentials for $t > 40$ fm$^{-2}$
give the result that is higher than the experimental data and the result with the Yamaguchi kernel.

The line denoted as YY (SSD4) stands for the $^1S_0$, $^3S_1$ and $^3D_1$ partial-wave states with $p_D=4\%$
calculated with the Yamaguchi potential. The contributions of the $P$ and $D$ partial-wave
state amplitudes were estimated by using the formulae for C and D~(\ref{4func2}),
and the result was found to be negligible.
An additional term in the function $\tau$, which corresponds to the $^3D_1$ partial-wave state
was also taken into account and the result was found to be noticeable.
Below, the SSD4 notation is used for calculations with the $^1S_0$, $^3S_1$ and $^3D_1$
partial-wave states with $p_D=4\%$ where the $^3D_1$ state
is taken into account only in function $\tau$.
{ The YY (SSD4) result is less than the YY (SS) one.}

The results for $p_D=5,6\%$ are very close to $p_D=4\%$ and are not plotted.
As one can see,
{ all considered rank-one potentials do not give diffraction minima in the form factor $\ffc$,
which exists in the experimental data.}

\begin{figure}[ht]
  \centering
  \includegraphics[width=0.85\linewidth,angle=0]{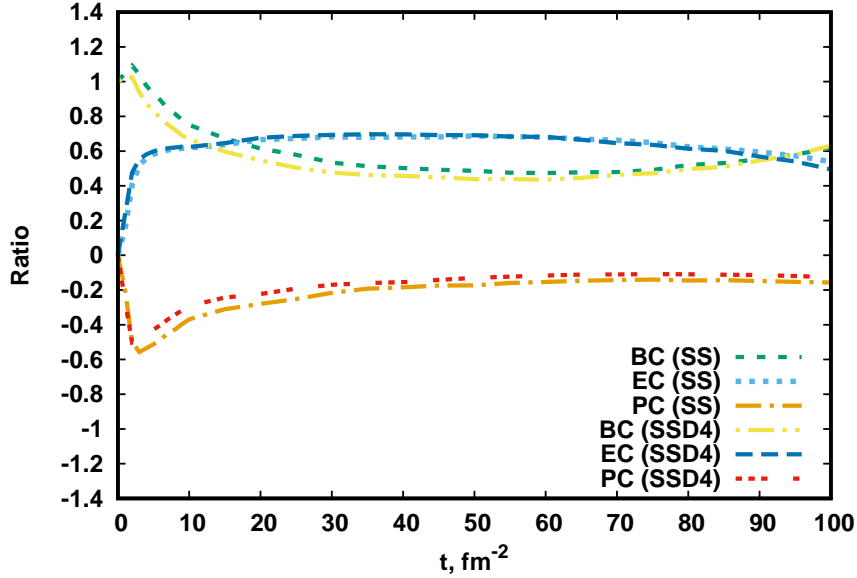}
  \caption{Ratio of the parts of relativistic corrections to their sum
    as a function of $t$ for the cases SS and SSD4.}
  \label{fig2}
\end{figure}

Figure~\ref{fig2} shows the ratios of three additive
parts of RCs to their sums for two Yamaguchi potentials without (SS) and
the $^3D_1$ partial-wave state (SSD4).
At high $t$ the contributions are rather flat.
{One can see that BC and EC are positive while PC is negative
and compensate about 20\% of BC+EC.
For example, for SS calculations
at t=50~fm$^{-2}$ the EC contribution is about 70\%,
BC is about 50\% and PC is about -20\%,
at t=100~fm$^{-2}$ the EC contribution is about 58\%,
BC is about 60\% and PC is about -18\%.}
The results for SSD4 are very similar especially at high momentum transfer squared.

\begin{figure}[ht]
  \centering
  \includegraphics[width=0.85\linewidth,angle=0]{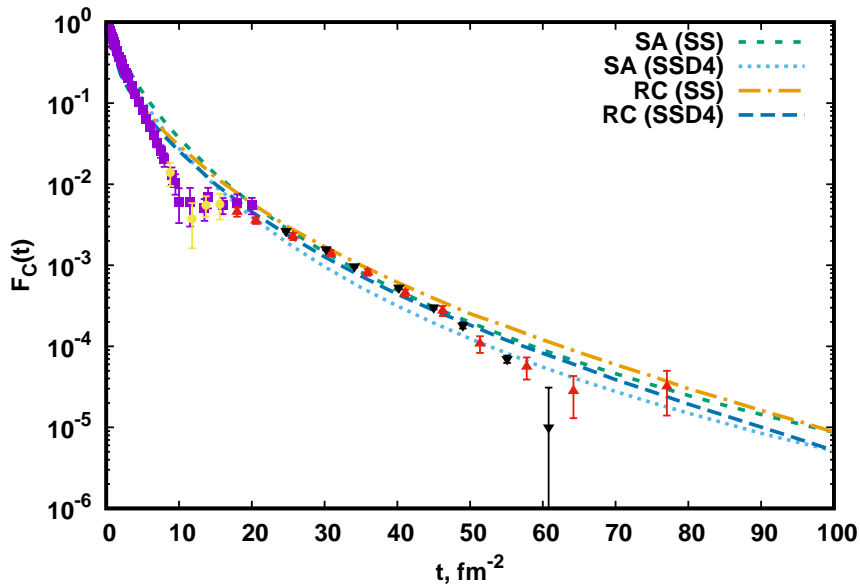}
  \caption{Static approximation and contribution of the relativistic corrections for the charge form factor
    of $^3$He as a function of $t$ for the cases SS and SSD4.
    The experimental data are the same as in Fig.~\ref{fig1}.}
  \label{fig3}
\end{figure}

Figure~\ref{fig3} shows the static approximation and the full relativistic
corrections (BC + PC + EC) for two cases: the $^1S_0$, $^3S_1$ (SS)
and the $^1S_0$, $^3S_1$ and $^3D_1$ partial-wave states (SSD4).
{For the SS case RC are higher than SA in the region 20-100 fm$^{-2}$
  by the maximum factor of 1.3 at t=50~fm$^{-2}$.
  For the SSD4 case RC are higher than SA in the region 15-100 fm$^{-2}$
  by the maximum factor of 1.5 at t=50~fm$^{-2}$.}

{
Taking into account the $D$ wave in the function $\tau$ (SSD4)
leads to the result that is less than calculation with only $S$ waves:}
{SA is lower by a factor of 0.7-0.65 at t=20-30~fm$^{-2}$ and
by a factor of 0.6 at t=50-100~fm$^{-2}$,
RC is lower by a factor of 0.85-0.8 at t=20-30~fm$^{-2}$ and
by a factor of 0.7-0.6 at t=50-100~fm$^{-2}$.}

It should be stressed that neither the inclusion of the $^3D_1$ partial-wave state nor
the relativistic corrections lead to zero in the $\het$ charge form factor.
{ Since the simple rank-one separable kernel considered in the paper does not reproduce the
diffraction minima of the form factor $\ffc$, one need to consider a multirank kernel
such as, for example in~\cite{Rupp:1991th}. In that paper, the first diffraction minimum
does appear but at an incorrect momentum transfer. Therefore, one needs
to investigate the influence of the relativistic corrections discussed in this paper
to the difraction
minima position. This work is in progress.
However, other effects such as relativistic interaction currents,
the off-shell behaviour of the EM nucleon form factors and the precise fitting of the separable
functions of the NN interaction can contribute as well.
}

\section{Summary}

In this paper, solutions of the BSF equation for $\het$ have been used to
calculate the charge form factor. Expressions for the form factor
were obtained by a straightforward relativistic generalization of
nonrelativistic expressions.
Multiple integration was performed by means of the Monte Carlo algorithm.
Moreover, in the calculations, the Lorentz transformation of the arguments
of the propagator and the final $\het$ wave function were taken into account.

Solutions with rank-one Yamaguchi potentials not only with $S$ two-nucleon
partial-wave states but also with $P$ and $D$ partial-wave states were taken into account.
The only inclusion of the $D$ wave into the function $\tau$
was found to be noticeable while the contributions of the $P$ and $D$ waves
by means of the amplitudes were found to be negligible.

Two approximations have been considered: static approximation and relativistic
corrections. It was found that the main contribution to the corrections
came from the Lorentz boost of the final $\het$ wave function arguments.
{At high momentum transfer squared,
  the relativistic corrections play a significant role
and are larger than the static approximation by a factor 1.3-1.5.}

\bibliographystyle{elsarticle-num}

\end{document}